%% file: CPC-2022-0124.tex
\documentclass[a4paper,10pt,twoside]{cpc-hepnp}

\usepackage{xcolor}

\usepackage{url}
\usepackage{multicol}
\usepackage{graphicx}
\usepackage{booktabs}
\usepackage{amssymb,bm,mathrsfs,bbm,amscd}
\usepackage[tbtags]{amsmath}
\usepackage{lastpage}
\usepackage{CJK}
\usepackage{lineno}
\usepackage{hyperref}

\usepackage[abs]{overpic}

\makeatletter
    
    \newcommand{\Rmnum}[1]{\expandafter\@slowromancap\romannumeral #1@}
\makeatother

\begin{document}

\begin{CJK*}{GBK}{song}

\footnotetext[0]{Submitted to Chinese Physics C}

\title{Search for a lighter neutral custodial fiveplet scalar in the Georgi-Machacek model
\thanks{Supported by National Natural Science Foundation of China (11875275, 12061141003, 11661141007),
China Ministry of Science and Technology (2018YFA0403901) and partially by the France China Particle
Physics Laboratory (FCPPL) and CAS Center for Excellence in Particle Physics (CCEPP)}}

\author{
Chu Wang$^{1,2}$%
\quad Jun-Quan Tao$^{1;1)}$\email{taojq@mail.ihep.ac.cn}%
\quad M. Aamir Shahzad$^{1,2}$%
\quad Guo-Ming Chen$^{1,2}$%
\quad S. Gascon-Shotkin$^{3}$%
\\
}

\maketitle

\address{
$^1$ Institute of High Energy Physics, Chinese Academy of Sciences, Beijing 100049, China \\
\vspace*{4pt}
$^2$ University of Chinese Academy of Sciences, Beijing 100049, China \\
$^3$ Institut de Physique des 2 Infinis de Lyon, Universit\'{e} de Lyon, Universit\'{e} Claude Bernard Lyon 1, CNRS-IN2P3,\\
     Villeurbanne 69622, France \\
}

\begin{abstract}
Many researches from both the theoretical and experimental
sides have been performed to search for a new Higgs Boson lighter than the 125~$GeV$ Higgs boson which was discovered at the LHC in 2012.
In this paper we explore the possibility of constraining a lighter neutral custodial fiveplet scalar $H_{5}^{0}$ in the Georgi-Machacek (GM) model by the latest results of the search for a lighter Higgs boson decaying into two photons from LHC data.
The custodial-singlet mass eigenstate $h$ or $H$ is considered to be the LHC observed 125~$GeV$ Higgs boson.
A new set of constrained parameters that is favoured by low-mass $H_{5}^{0}$, is proposed to generate events efficiently.
The production of $H_{5}^{0}$ from the scan based on the constrained parameters is compared to the latest results of the search for a lighter Higgs boson decaying into two photons by CMS Collaboration, after the theoretical constraints from GM model and the constraints from all existing relevant experimental measurements including the recent results of the Higgs boson searches from the LHC.
The numerical analyses of the surviving GM parameter space are performed.
The tendencies and correlations of the GM input parameters from the phenomenological studies are summarized.
In addition the discovery potential of other interesting decay channels of this low-mass neutral custodial fiveplet scalar are discussed.
\end{abstract}


\begin{multicols}{2}

\section{Introduction}

The standard model (SM) of particle physics ~\cite{Glashow:1961tr,Weinberg:1967tq,Salam:1968rm} can explain the high-energy experimental results successfully.
Particle masses arise from the spontaneous breaking of electroweak symmetry, which is achieved through the Brout-Englert-Higgs (BEH) mechanism~\cite{Englert:1964et,Higgs:1964ia,Higgs:1964pj,Guralnik:1964eu,Kibble:1967sv,Higgs:1966ev}.
In this BEH mechanism only one scalar field remains with its corresponding quantum, the Higgs boson, finally was discovered at the LHC with its mass approximately
125~GeV~\cite{Aad:2012tfa,Chatrchyan:2012xdj,Aad:2013wqa,Chatrchyan:2013lba}.
The latest measurements~\cite{Aad:2019mbh,Sirunyan:2020xwk,Sirunyan:2019twz,Sirunyan:2018koj,Sirunyan:2018sgc} of this Higgs boson at the LHC show no bias from the SM predicted Higgs boson. However, many important questions about the nature and the origin of the Higgs boson
discovered at the LHC are still not answered.
SM can not explain many observations, such as dark matter and neutrino masses,  and many puzzles including the
hierarchy problem and the strong-CP problem~\cite{Patrignani:2016xqp}.
Some physics beyond the SM (BSM) can provide a Higgs boson that is compatible with the observed 125~GeV Higgs boson
and can address some of the questions that are left unanswered by the SM.
The extended Higgs sector of these BSM models, for example the Next-to-Minimal Supersymmetric Standard Model (NMSSM)~\cite{Ellwanger:2009dp,Cao:2012fz} and the generalized Two Higgs Doublets Model (2HDM)~\cite{Celis:2013rcs,Chang:2013ona}
, can also provide additional Higgs bosons with masses below 125~GeV which can give rise to
a rich and interesting phenomenology~\cite{Jia-Wei:2013eea,Ellwanger:2015uaz,Tao:2018zkx,Cacciapaglia:2016tlr}.

One more phenomenologically interesting model is the Georgi-Machacek (GM) model~\cite{Georgi:1985nv,Chanowitz:1985ug}
which provides a prototype for extensions of the SM Higgs sector by adding scalars in isospin-triplet to preserve custodial SU(2) symmetry.
The GM model can be generalized to include scalars in isospin representations larger than triplets under the custodial symmetry~\cite{Logan:2015xpa}.
The physical fields of the generalized GM model can be transformed as a fiveplet, a triplet and two singlets.
The couplings of the SM-like Higgs boson in the GM model to $WW$ and to $ZZ$ can be larger than these in the SM.
The singly- and doubly-charged scalars couple at tree level to vector boson pairs.
These futures make this model interesting to the experimental communities to perform direct searches for example at the LHC~\cite{ATLAS:2014kca,Aaboud:2018qcu,Khachatryan:2014sta,Sirunyan:2017sbn}. The LHC experiments have sensitivity to the production
of the fermiophobic custodial-fiveplet states, $H_{5}^{0}$, $H_{5}^{\pm}$ and $H_{5}^{\pm\pm}$.
A lot of studies with their masses ($m_{5}$) lager than 200~GeV have been performed, as guided with so-called H5-plane benchmark by the LHC Higgs Cross Section Working Group~\cite{LHCHiggsCrossSectionWorkingGroup:2016ypw} with the detailed phenomenological studies in~\cite{Logan:2017jpr}.
Both ATLAS and CMS have performed the searches for diphoton resonances in the mass regions lower than 125~GeV.
It's interesting to check the phenomenology of the neutral custodial fiveplet scalar $H_{5}^{0}$ in the lower mass range.
Recently a new benchmark plane so called low-$m_{5}$ benchmark was introduced and studied~\cite{Ismail:2020kqz}.
For the study of updated constraints on the GM model from LHC Run2 with the detail in~\cite{Ismail:2020zoz},
the ATLAS results of the diphoton resonance search with LHC Run1 8~TeV data~\cite{Aad:2014ioa} was used to
compare with the GM $H_{5}^{0}\rightarrow\gamma\gamma$ decays from the general scan with 10,000 points.
CMS Collaboration published the results of the search for low-mass Higgs bosons in the mass range from 70~$GeV$ to 110~$GeV$ in the diphoton channel with the full 2016 dataset
at $\sqrt{s}$ = 13 TeV~\cite{Sirunyan:2018aui}, which show more sensitivity in the same mass range than the ATLAS result with 80$~fb^{-1}$ data at $\sqrt{s}$ = 13 TeV~\cite{ATLAS:2018xad} as illustrated in Figure 1 in Ref.~\cite{Heinemeyer:2018wzl}.
So the CMS searching results have more stringent constraint on the GM model.

In this paper the phase space of the free parameters of the GM model which is more favored for the low-mass neutral custodial fiveplet scalar $H_{5}^{0}$, is studied carefully.
We explore the possibility of constraining the low-mass $H_{5}^{0}$ in the GM model
by comparing the production rates of the $H_{5}^{0}\rightarrow\gamma\gamma$ decays with the latest CMS results\cite{Sirunyan:2018aui}, after the theoretical constraints and the constraints from the experimental measurements.
The phenomenology of the surviving GM parameter space from this comparison is summarized.
In addition the discovery potential of other interesting decay channels of this low-mass neutral custodial fiveplet scalar are studied and discussed.
The structure of this paper is organized as follows. In section 2,
we introduce the GM model briefly and
the parameter ranges we choose for the scan.
The numerical analyses and results are described in section 3.
Finally the conclusions are presented in section 4.

\section{GM model and constraints on GM} \label{sec:GM}

\subsection{Brief description of GM model}

The scalar sector of the GM model~\cite{Georgi:1985nv,Chanowitz:1985ug} consists of the usual complex doublet ($\phi^+$, $\phi^0$)
, a real triplet ($\xi^+$, $\xi^0$, $\xi^-$)
and a complex triplet ($\chi^+$, $\chi^+$, $\chi^0$).
In order to make the global SU(2)$_L \times$SU(2)$_R$ symmetry explicit, the doublet is written in the form of a bi-doublet $\Phi$ while the triplets are combined to form a bi-triplet $X$:
\begin{equation}
        \Phi = \left( \begin{array}{cc}
        \phi^{0*} &\phi^+  \\
        -\phi^{+*} & \phi^0  \end{array} \right), \qquad
        X =
        \left(
        \begin{array}{ccc}
        \chi^{0*} & \xi^+ & \chi^{++} \\
         -\chi^{+*} & \xi^{0} & \chi^+ \\
         \chi^{++*} & -\xi^{+*} & \chi^0
        \end{array}
        \right).
        \label{eq:BiDBiTPX}
\end{equation}
The vevs (vacuum expectation values) are defined by $\langle \Phi  \rangle = \frac{ v_{\phi}}{\sqrt{2}} I_{2\times2}$  and $\langle X \rangle = v_{\chi} I_{3 \times 3}$, where $I_{2\times2}$ and $I_{3\times3}$ are the unit matrix.
$W$ and $Z$ boson masses constrain
\begin{equation}
        v_{\phi}^2 + 8 v_{\chi}^2 \equiv v^2 = \frac{1}{\sqrt{2} G_F} \approx (246~{\rm GeV})^2
        \label{eq:vevrelation}
\end{equation}
with $G_F$ being the Fermi constant.

The most general gauge-invariant scalar potential involving these fields that conserves custodial SU(2) is given by
\begin{eqnarray}
        V(\Phi,X) &= & \frac{\mu_2^2}{2}  \text{Tr}(\Phi^\dagger \Phi)
        +  \frac{\mu_3^2}{2}  \text{Tr}(X^\dagger X) \nonumber \\
        & & + \lambda_1 [\text{Tr}(\Phi^\dagger \Phi)]^2
        + \lambda_2 \text{Tr}(\Phi^\dagger \Phi) \text{Tr}(X^\dagger X)   \nonumber \\
    & & + \lambda_3 \text{Tr}(X^\dagger X X^\dagger X)
    + \lambda_4 [\text{Tr}(X^\dagger X)]^2   \nonumber \\
    & & - \lambda_5 \text{Tr}( \Phi^\dagger \tau^a \Phi \tau^b) \text{Tr}( X^\dagger T^a_1 X T^b_1) \nonumber \\
    & &  - M_1 \text{Tr}(\Phi^\dagger \tau^a \Phi \tau^b)(U X U^\dagger)_{ab} \nonumber \\
    & & -  M_2 \text{Tr}(X^\dagger T^a X T^b)(U X U^\dagger)_{ab}.
    \label{eq:potential}
\end{eqnarray}
Here the SU(2) generators for the doublet representation are $\tau^a = \sigma^a/2$ with $\sigma^a$ being the Pauli matrices and the generators for the triplet representation $T^a_1$ are
\begin{equation}
\begin{split}
T^1 =
\left(
\begin{array}{ccc}
 0 & \frac{1}{\sqrt{2}} & 0 \\
 \frac{1}{\sqrt{2}} & 0 & \frac{1}{\sqrt{2}} \\
 0 & \frac{1}{\sqrt{2}} & 0 \\
\end{array}
\right), \qquad
T^2 =
\left(
\begin{array}{ccc}
 0 & -\frac{i}{\sqrt{2}} & 0 \\
 \frac{i}{\sqrt{2}} & 0 & -\frac{i}{\sqrt{2}} \\
 0 & \frac{i}{\sqrt{2}} & 0 \\
\end{array}
\right), 
\\
T^3 =
\left(
\begin{array}{ccc}
 1 & 0 & 0 \\
 0 & 0 & 0 \\
 0 & 0 & -1 \\
\end{array}
\right).
\end{split}
\end{equation}
The matrix $U$
is given by~\cite{Aoki:2007ah}
\begin{equation}
         U = \left( \begin{array}{ccc}
        - \frac{1}{\sqrt{2}} & 0 &  \frac{1}{\sqrt{2}} \\
         - \frac{i}{\sqrt{2}} & 0  &   - \frac{i}{\sqrt{2}} \\
           0 & 1 & 0 \end{array} \right).
         \label{eq:U}
\end{equation}

The physical fields can be organized by their transformation properties under the custodial SU(2) symmetry into two singlets, a triplet and a fiveplet.
The triplet and fiveplet states are given by
\begin{eqnarray}
        &&H_3^+ = - s_H \phi^+ + c_H \frac{\left(\chi^++\xi^+\right)}{\sqrt{2}}, \qquad
        H_3^0 = - s_H \phi^{0,i} + c_H \chi^{0,i}, \nonumber \\
        &&H_5^{++} = \chi^{++}, \qquad
        H_5^+ = \frac{\left(\chi^+ - \xi^+\right)}{\sqrt{2}}, \nonumber \\
        &&H_5^0 = -\sqrt{\frac{2}{3}} \xi^0 + \sqrt{\frac{1}{3}} \chi^{0,r},
\end{eqnarray}
where the vevs are parameterized by
\begin{equation}
        c_H \equiv \cos\theta_H = \frac{v_{\phi}}{v}, \qquad
        s_H \equiv \sin\theta_H = \frac{2\sqrt{2}\,v_\chi}{v},
\end{equation}
and the neutral fields have been decomposed into real and imaginary parts according to
\begin{eqnarray}
        &&\phi^0 \to \frac{v_{\phi}}{\sqrt{2}} + \frac{\phi^{0,r} + i \phi^{0,i}}{\sqrt{2}},
        \qquad
        \chi^0 \to v_{\chi} + \frac{\chi^{0,r} + i \chi^{0,i}}{\sqrt{2}},
        \nonumber \\
        &&\xi^0 \to v_{\chi} + \xi^0.
        \label{eq:decomposition}
\end{eqnarray}
The masses within each custodial multiplet are degenerate at tree level and can be, after eliminating $\mu_2^2$ and $\mu_3^2$ in favor of the vevs, written as\footnote{Note that the ratio $M_1/v_{\chi}$ is finite in the limit $v_{\chi} \to 0$,
        $\frac{M_1}{v_{\chi}} = \frac{4}{v_{\phi}^2}
        \left[ \mu_3^2 + (2 \lambda_2 - \lambda_5) v_{\phi}^2
        + 4(\lambda_3 + 3 \lambda_4) v_{\chi}^2 - 6 M_2 v_{\chi} \right]$,
which follows from the minimization condition $\partial V/\partial v_{\chi} = 0$.}
\begin{eqnarray}
        m_5^2 &=& \frac{M_1}{4 v_{\chi}} v_\phi^2 + 12 M_2 v_{\chi}
        + \frac{3}{2} \lambda_5 v_{\phi}^2 + 8 \lambda_3 v_{\chi}^2, \nonumber \\
        m_3^2 &=&  \frac{M_1}{4 v_{\chi}} (v_\phi^2 + 8 v_{\chi}^2)
        + \frac{\lambda_5}{2} (v_{\phi}^2 + 8 v_{\chi}^2).
\end{eqnarray}
The two singlet mass eigenstates are given by
\begin{eqnarray}
        h &=& \cos \alpha \, \phi^{0,r} - \sin \alpha \, H_1^{0\prime},  \nonumber \\
        H &=& \sin \alpha \, \phi^{0,r} + \cos \alpha \, H_1^{0\prime},
\end{eqnarray}
where
\begin{equation}
        H_1^{0 \prime} = \sqrt{\frac{1}{3}} \xi^0 + \sqrt{\frac{2}{3}} \chi^{0,r}.
\end{equation}
The mixing angle and masses are given by
\begin{eqnarray}
        \sin 2 \alpha =  \frac{2 \mathcal{M}^2_{12}}{m_H^2 - m_h^2},    \qquad
        \cos 2 \alpha =  \frac{ \mathcal{M}^2_{22} - \mathcal{M}^2_{11}  }{m_H^2 - m_h^2},
        \nonumber \\
        m^2_{h,H} = \frac{1}{2} \left[ \mathcal{M}_{11}^2 + \mathcal{M}_{22}^2
        \mp \sqrt{\left( \mathcal{M}_{11}^2 - \mathcal{M}_{22}^2 \right)^2
        + 4 \left( \mathcal{M}_{12}^2 \right)^2} \right]. \nonumber \\
        \label{eq:hmass}
\end{eqnarray}
The elements of their mass matrix are given by
\begin{eqnarray}
        \mathcal{M}_{11}^2 &=& 8 \lambda_1 v_{\phi}^2, \nonumber \\
        \mathcal{M}_{12}^2 &=& \frac{\sqrt{3}}{2} v_{\phi}
        \left[ - M_1 + 4 \left(2 \lambda_2 - \lambda_5 \right) v_{\chi} \right], \nonumber \\
        \mathcal{M}_{22}^2 &=& \frac{M_1 v_{\phi}^2}{4 v_{\chi}} - 6 M_2 v_{\chi}
        + 8 \left( \lambda_3 + 3 \lambda_4 \right) v_{\chi}^2.
\end{eqnarray}

We define $H$ to be heavier than $h$ ($m_h < m_H$), and either $h$ or $H$ can be the observed 125~GeV Higgs boson at LHC.

\subsection{Constraints on GM and its parameters}  \label{sec:ConstraintAndPara}

The program package $GMCALC$ (version
1.5.0)~\cite{Hartling:2014xma} with Fortran code is employed in this study to calculate
the mass spectrum of the Higgs bosons in the GM model, their decaying branching ratios (BR) and total widths,
their relevant mixing angles, as well as the tree-level couplings
of the Higgs bosons to other particles.
It also includes a routine to generate the datacard "$param\_card.dat$"
to be used by $MadGraph5\_aMC@NLO$~\cite{Alwall:2014hca} with the corresponding
FeynRules~\cite{Degrande:2011ua} model implementation, for the cross sections of the GM Higgs bosons.
The FeynRules implementation for the GM model includes automatic calculation of the next-to-leading order QCD corrections.
The Universal FeynRules Output (UFO) file of the FeynRules implementation for the GM model,
which can be downloaded from~\cite{GMUFOFileCite}, is used by $MadGraph5\_aMC@NLO$ (version 2.7.2) in this study.
Theoretical constraints in this paper include the conditions for tree-level unitarity,
bounded-from-below requirement on the scalar potential and the absence of deeper custodial symmetry-breaking minima,
as detailed in~\cite{Hartling:2014xma}.
Indirect constraints from the $S$ parameter, flavor physics $b \rightarrow s\gamma$ and $B_{s}^{0} \rightarrow \mu^{+}\mu^{-}$,
are also considered.
Constraints from the public tools $HiggsBounds$-$5$~\cite{Bechtle:2020pkv} (version
5.8.0) and $HiggsSignals$-$2$~\cite{Bechtle:2020uwn} (version
2.5.1) are applied
to further compare the predictions on the custodial-singlet mass eigenstate $h$ or $H$ with
LHC Higgs search results of various channels from Run2 at a center-of-mass energy of 13 TeV
including the cross section limits, signal rate and mass measurements, as well as the results in the form of simplified template cross section measurements.

A new low-$m_{5}$ benchmark for the GM model, defined for the neutral custodial fiveplet scalar with its mass $m_{5}$ $\in$ (50, 550) GeV,
was proposed in~\cite{Ismail:2020kqz} to study the phenomenological behavior
of the $H_{5}$ states and the SM-like Higgs, $h$.
In this reference~\cite{Ismail:2020kqz}, $\lambda_{3}$ (= -1.5), $\lambda_{4}$ (= 1.5 = -$\lambda_{3}$)and $M_{2}$ (= 20 GeV) were fixed to be constants,
together with the lightest custodial-singlet mass eigenstate $h$ to be the LHC observed 125 GeV Higgs boson.
In addition, the parameters $\lambda_{2}$ and $\lambda_{5}$ were fixed to be as functions of the mass $m_{5}$, $\lambda_{2}$ = 0.08($m_{5}$/100 GeV) and $\lambda_{5}$ = -0.32($m_{5}$/100 GeV)= -4$\lambda_{2}$.
Only $m_{5}$ and $s_{H}$ were allowed to float in the studied ranges, $m_{5}$ $\in$ (65, 550) GeV and $s_{H}$ $\in$ (0, 1).
After several iterations of tests, in order to generate the points efficiently we employ the following ranges  for the six specific parameters
\begin{equation}
\label{eq:gmpara}
\begin{split}
0.0 < M_{2} < 50, \quad 0.0 < s_{H} < 0.66, \quad -1.6 < \lambda_{3} < 0.0, \\
0.0 < \lambda_{4} < 1.6, \quad 0.0 < \lambda_{2} < 0.08(m_{5}/50 GeV), \\
-0.32(m_{5}/50 GeV) < \lambda_{5} < 0.0.
\end{split}
\end{equation}
We find that wider ranges of these parameters have practically no impact on our conclusions.
Same as ~\cite{Ismail:2020kqz}, the Fermi constant $G_{F}$ is set to be $1.1663787 \times 10^{-5}$ $GeV^{-2}$.
In order to study more stringent constraint on the GM model by using the results of the search for low-mass Higgs bosons in the mass range from 70~$GeV$ to 110~$GeV$ in the diphoton channel at CMS ~\cite{Sirunyan:2018aui} and to study the discovery potential of other interesting decay channels, the initial scan range of the $H_{5}^{0}$ mass are specified in the range $m_{5}$ $\in$ (65, 130) GeV.
In addition, any of the custodial-singlet mass eigenstate $h$ or $H$ could be the LHC observed Higgs boson,
so should pass the constraints of $HiggsBounds$-$5$ and $HiggsSignals$-$2$ as mentioned in above paragraph.

\section{Numerical analyses} \label{sec:h5study}

In this study we focus on the phenomenological behaviors
of the neutral custodial fiveplet scalar state $H_{5}^{0}$ decaying into two photons and some other interesting final states.
After implementations of the constraints as mentioned in previous subsection 2.2,
production of $H_{5}^{0}$ from the scan is compared to
the results of the search for low-mass Higgs bosons in the mass range from 70~$GeV$ to 110~$GeV$ in the diphoton channel with the full 2016 dataset
at $\sqrt{s}$ = 13 TeV with the CMS experiment at the LHC~\cite{Sirunyan:2018aui}.
The favored GM parameter phase space from this comparison is summarized.
The discovery potential of other interesting decay channels of this low-mass neutral custodial fiveplet scalar is also studied and discussed.

\subsection{Cross section} \label{sec:h5xs}

The cross sections of the Drell-Yan production of $H_{5}^{0}H_{5}^{+}$ and $H_{5}^{0}H_{5}^{-}$ separately at next-to-leading order (NLO) in QCD for 13 TeV pp collisions are firstly generated using $MadGraph5\_aMC@NLO$ (version 2.7.2),
together with the UFO file downloaded from~\cite{GMUFOFileCite} for GM model and the datacard "$param\_card.dat$" generated with $GMCALC$ (version
1.5.0). The PDF4LHC15 NLO parton distribution
functions (PDF)~\cite{Butterworth:2015oua}, $PDF4LHC15\_nlo\_30\_pdfas$, are employed.
In this PDF, 30 sets of symmetric eigenvectors give the PDF systematic uncertainties,
and 2 additional sets with $\alpha_{s}(M_{Z})$ = 0.1165 and $\alpha_{s}(M_{Z})$ = 0.1195 separately which are different from the central value ($\alpha_{s}(M_{Z})$ = 0.118) are used to estimate the systematic uncertainties from $\alpha_{s}$.
Typically the combined PDF + $\alpha_{s}$ uncertainty is about 2.0\% on the cross sections for both $H_{5}^{0}H_{5}^{+}$ and $H_{5}^{0}H_{5}^{-}$.
For the uncertainties from QCD scales of renormalization ($\mu_{R}$) and factorization scales ($\mu_{F}$),
practically they are obtained from the cross section results of nine scale configurations by combining $\mu_{R}$/$\mu_{0}$ = (0.5, 1, 2) and $\mu_{F}$/$\mu_{0}$ = (0.5, 1, 2) independently, with the central value of the scale factor $\mu_{0}$ to be the sum of the transverse masses divided by two of all final state particles and partons.
Then the maximum value and minimum value among these results are used to evaluate the scale uncertainties.
Typically the scale uncertainty varies from about 1.5\% at $m_{5}$ = 130 GeV to about 4.0\% at $m_{5}$ = 65 GeV.
Fig.~\ref{fig:h5XS} shows the total cross sections of these two processes as a function of $H_{5}^{0}$ mass in picobarns (pb), with the total uncertainties from PDF, $\alpha_{s}$ and QCD scales included as the filled area around the central values.
The Drell-Yan production cross sections of $H_{5}^{0}H_{5}^{+}$ and $H_{5}^{0}H_{5}^{-}$ are independent of the six specific parameters in formula~\ref{eq:gmpara}, but are dependent on the mass $m_{5}$.

\begin{center}
\includegraphics[width=0.4\textwidth,height=0.4\textwidth]{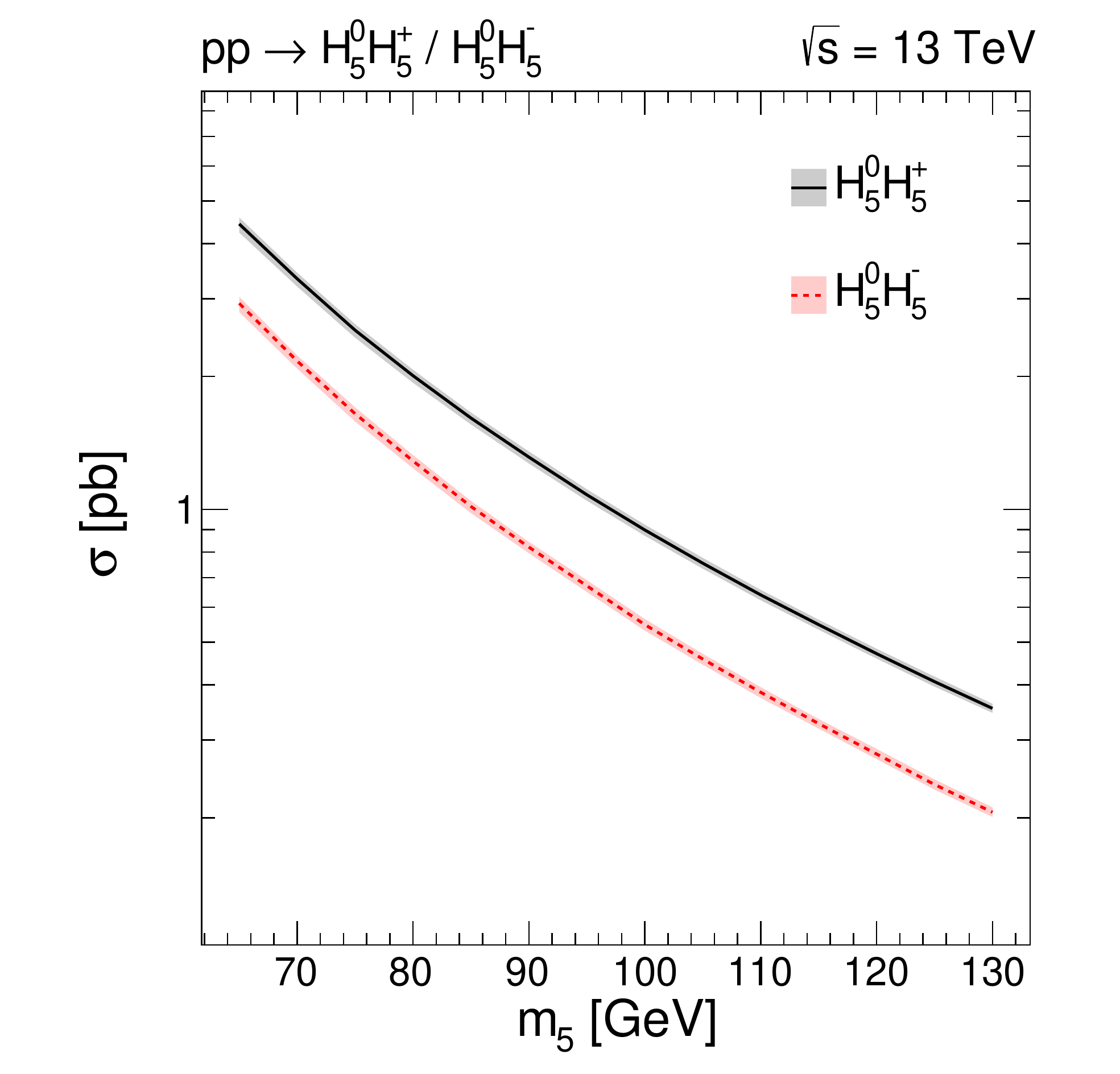}
\figcaption{\label{fig:h5XS} (color online) Cross sections of $H_{5}^{0}H_{5}^{+}$ and $H_{5}^{0}H_{5}^{-}$ at the LHC with
$\sqrt{s}$ = 13 TeV as a function of $H_{5}^{0}$ mass ($m_{5}$) in picobarns (pb), computed using $MadGraph5\_aMC@NLO$ at NLO in QCD, with the total uncertainties from PDF, $\alpha_{s}$ and QCD scales shown as the filled area around the central values with lines. }
\end{center}

\subsection{Scan results with constrained parameters} \label{sec:h5scan}

In this subsection, we will explore the possibility that the
signal may be given by the neutral custodial fiveplet scalar state $H_{5}^{0}$
in the GM model. We firstly performed the comparison of
the production cross sections of $H_{5}^{0}\rightarrow\gamma\gamma$ from the GM scans with constrained parameters as introduced in subsection 2.2
and the CMS observed upper limit of the production cross sections with the full 2016 data set at $\sqrt{s}$ = 13 TeV~\cite{Sirunyan:2018aui}.
Then the six specific parameters were detailed studied by applying the CMS searching results.
In this study, a general scan of about six million
points that satisfied all theoretical constraints and the indirect constraints from the S parameter and from flavor physics as described in subsection 2.2,
were randomly generated in the parameter phase space as specified in formula~\ref{eq:gmpara} and mass region $m_{5}$ $\in$ (65, 130) GeV.
With $HiggsBounds$-$5$, the predictions of either $h$ or $H$ from the scan should be consistent with the search results of the LHC observed Higgs boson in various channels from Run2 at a center-of-mass energy of 13 TeV. With $HiggsSignals$-$2$, the compatibility of $h$ or $H$  from the scans with the LHC measured results of the signal strengths
and mass of Higgs boson in Run2 by evaluating a $\chi^{2}$ calculation then its associated p-value, is also checked.
For scan points compatible with
the experimental constraints, the p-value given by
$HiggsSignals$-$2$ should be greater than 0.05.
By applying the constraints from these two programs,
about 5.5 million points are left for further analysis in the following paragraph and subsections.

The production rates in pb of $H_{5}^{0}$ decaying into $\gamma\gamma$, ($\sigma$ $\times$ BR)$_{H_{5}^{0}\rightarrow\gamma\gamma}$ with $\sigma$ the cross section of $H_{5}^{0}$ production and $BR$ the branching ratio of $H_{5}^{0}$ decaying into $\gamma\gamma$, versus $H_{5}^{0}$ mass are shown in Fig.~\ref{fig:h5ggXS}.
The observed exclusions or upper limits of the CMS Collaboration with
2016 data set at $\sqrt{s}$ = 13 TeV~\cite{Sirunyan:2018aui} is superimposed in the plot, as shown in red line.
It can be seen that many of the points, above the CMS observed exclusions can be excluded in the mass region $m_{5}$ $\in$ (70, 110) GeV at the 95\% confidence level (CL).
But there are still a lot of points with lower production rates could not be excluded by the CMS observed upper limits.

\begin{center}
        \includegraphics[width=0.4\textwidth,height=0.4\textwidth]{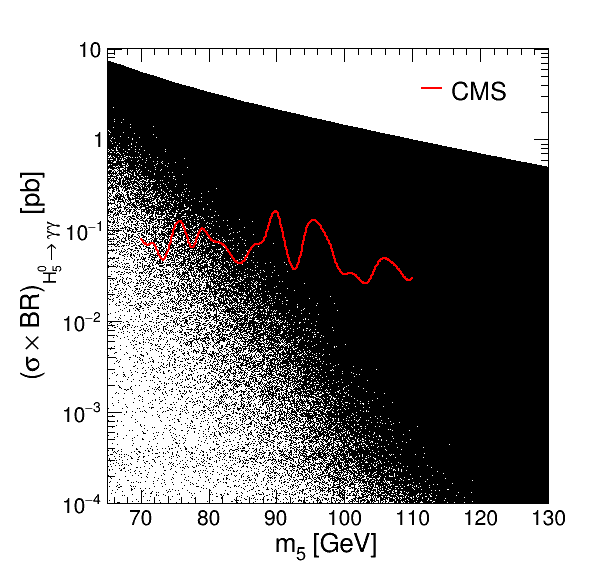}
        \figcaption{\label{fig:h5ggXS}
        (color online) Cross sections times the branching ratio of $H_{5}^{0}\rightarrow\gamma\gamma$ from the scan (black points) with the latest CMS observed exclusions~\cite{Sirunyan:2018aui} (red line) superimposed for comparison. }
\end{center}

Thanks to this CMS analysis~\cite{Sirunyan:2018aui}, we can try to constrain
the GM parameter spaces by checking the parameter distributions of the points below the CMS observed upper limits.
Fig.~\ref{fig:1dScan} show the comparisons between the distribution of all selected points (black histogram) and the distribution of the points that are not excluded by the CMS observe upper limits (red filled histogram) on top of the constraints as described in subsection 2.2, for each parameter of $\lambda_2$, $\lambda_3$, $\lambda_4$, $\lambda_5$, $M_2$ and $sin{\theta_H}$.
Both distributions are normalized to unity. One can find that the CMS observation could exclude some regions for some parameters, such as $\lambda_3$ and $\lambda_4$.
For these points which are not excluded by the CMS observed upper limits, $\lambda_3$ apparently shows a tendency to higher values with the points peaked at around -0.1.
While $\lambda_4$ favors lower values with a peak at around 0.6, showing an opposite behavior compared to $\lambda_3$.
One can also observe that $M_2$ prefers lower mass values while $sin{\theta_H}$ tends to accumulate at the middle of the scanned range with a peak at around 0.45.
For the other two parameters, $\lambda_2$ and $\lambda_5$, there are not very clear tendencies in the scanned ranges.

\begin{center}

        \includegraphics[width=0.23\textwidth,height=0.23\textwidth]{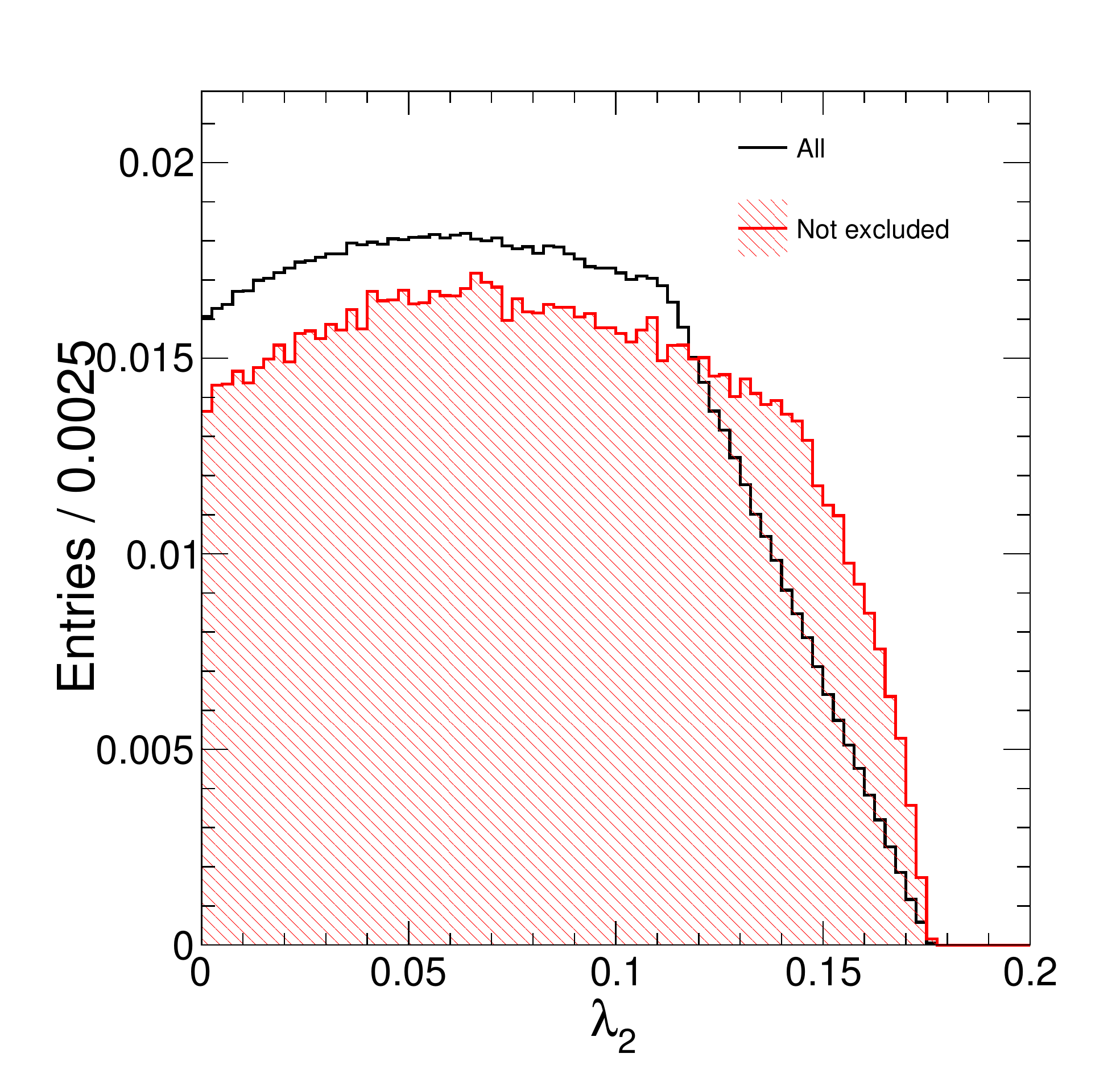}
        \includegraphics[width=0.23\textwidth,height=0.23\textwidth]{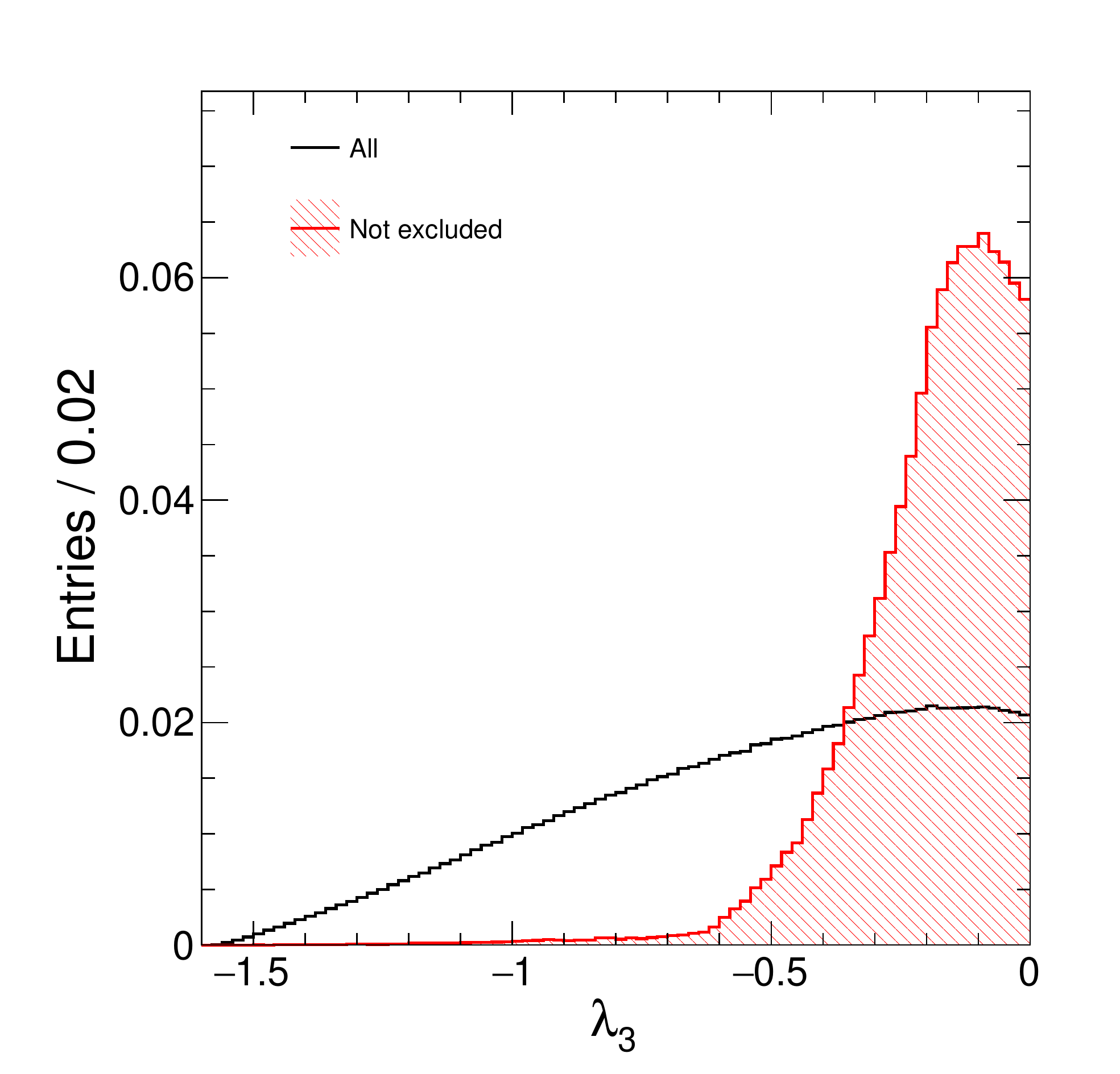} \\
        \includegraphics[width=0.23\textwidth,height=0.23\textwidth]{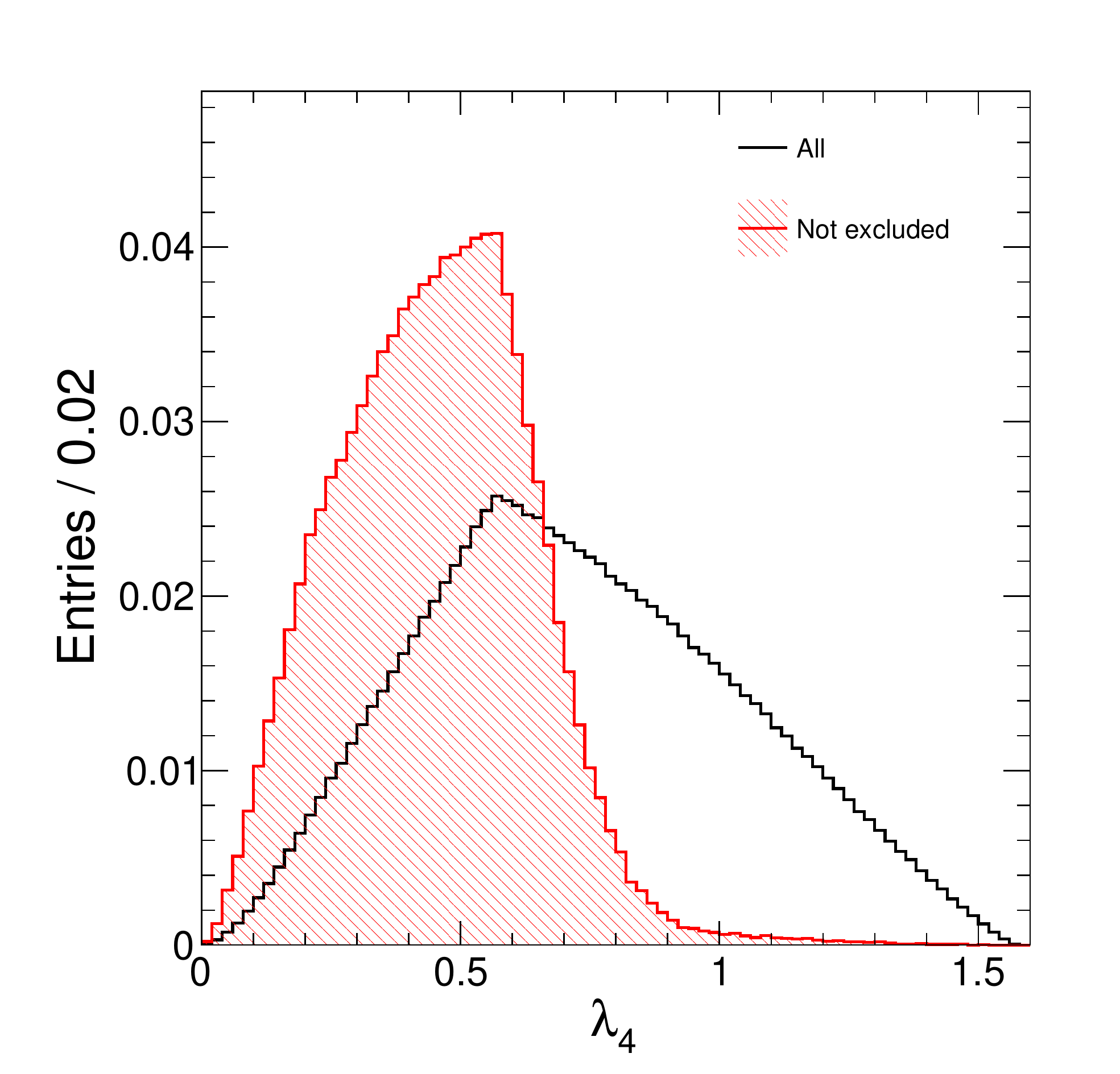}
        \includegraphics[width=0.23\textwidth,height=0.23\textwidth]{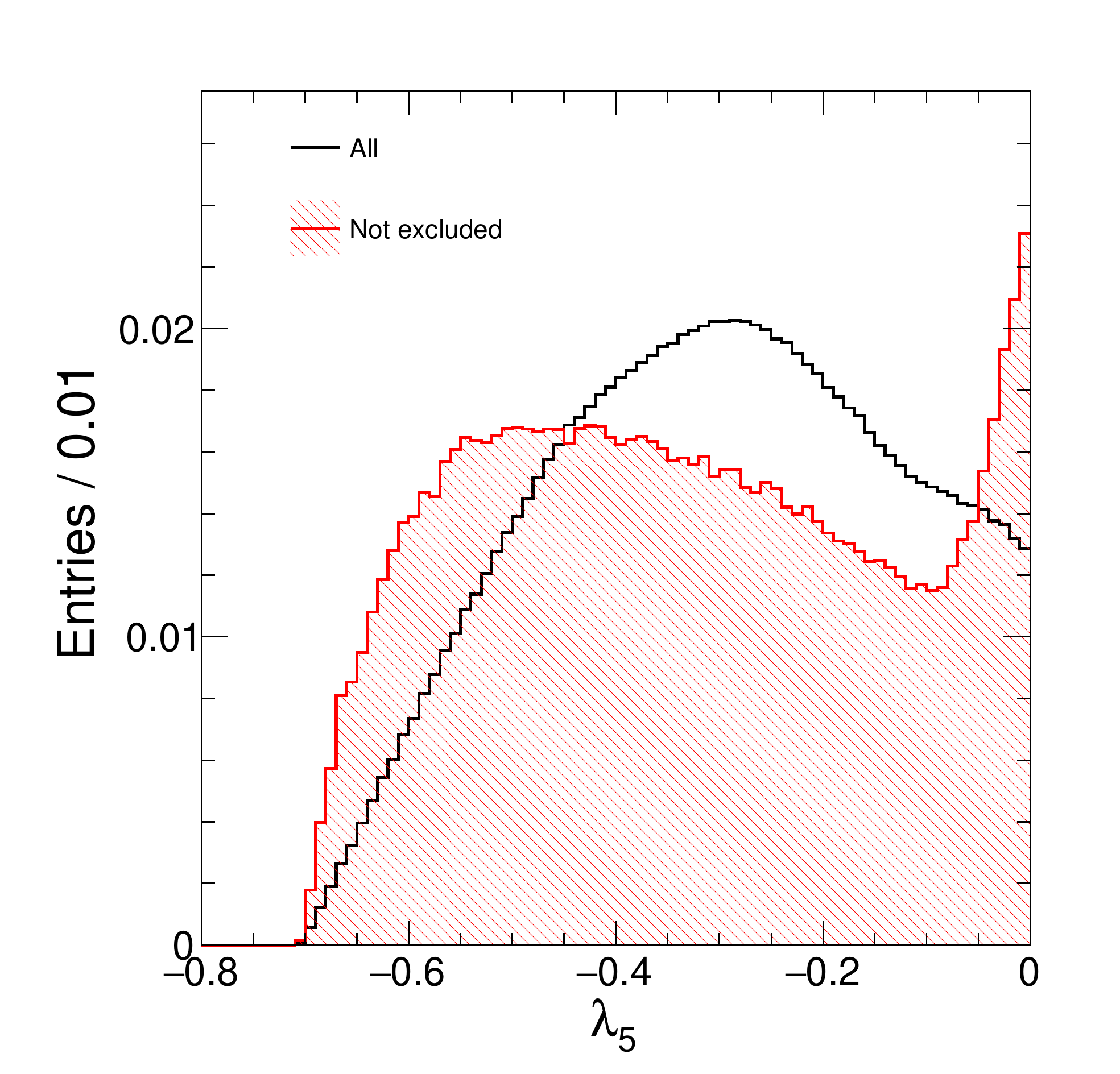} \\
        \includegraphics[width=0.23\textwidth,height=0.23\textwidth]{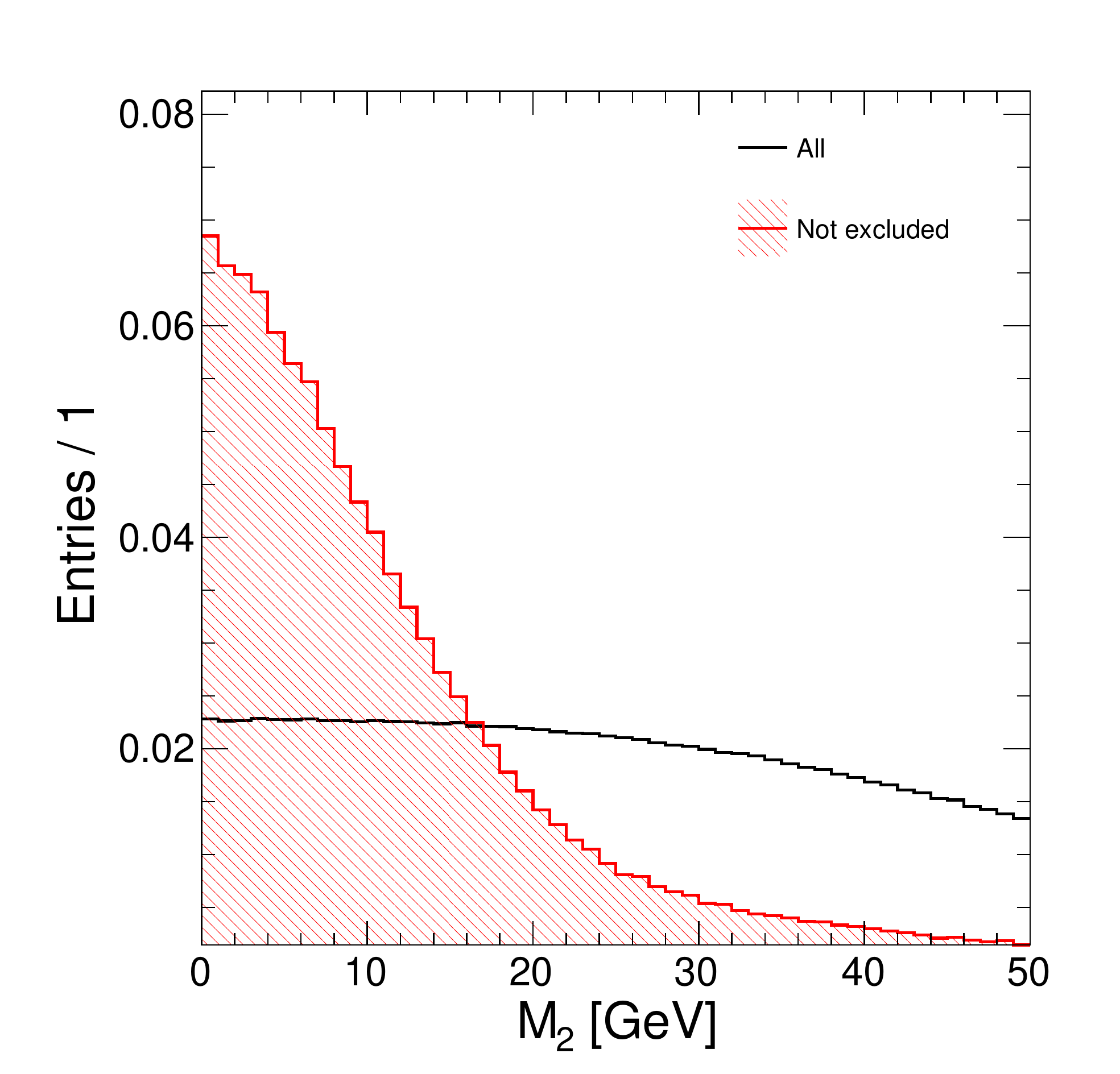}
        \includegraphics[width=0.23\textwidth,height=0.23\textwidth]{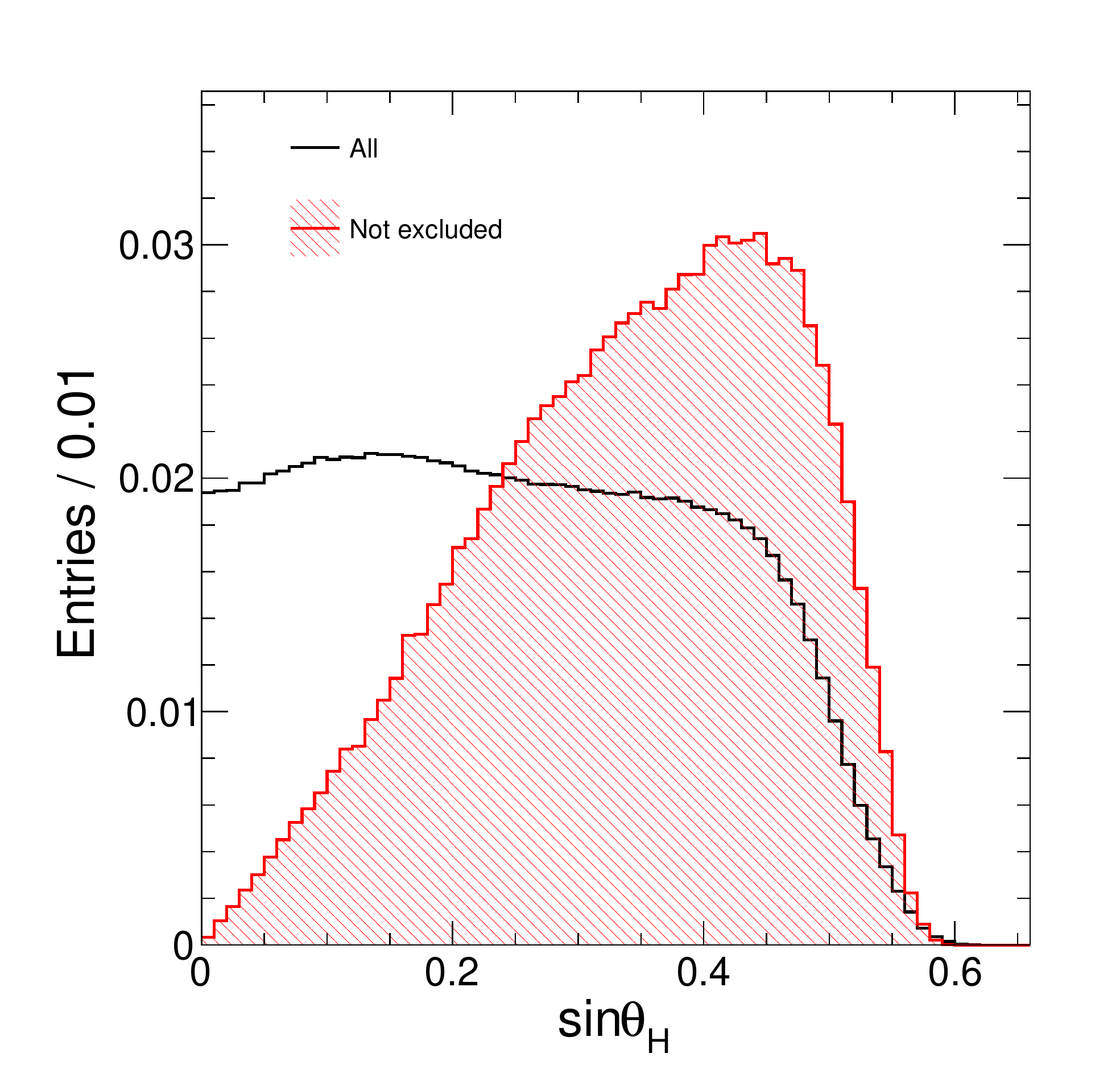}
        \figcaption{\label{fig:1dScan} (color online) Comparisons between the distribution of all selected points (black histogram) after the constraints and the distribution of the points that are not excluded by the CMS observed upper limits (red filled histogram) on top of the constraints, for each parameter of $\lambda_2$, $\lambda_3$, $\lambda_4$, $\lambda_5$, $M_2$ and $sin{\theta_H}$. Both distributions are normalized to unity.
        }
\end{center}

\subsection{Correlations between GM parameters} \label{sec:2Dscan}

We also explore the correlations between the scanned parameters
of the scan points passing all the constraints as described in subsection 2.2 and surviving from the CMS exclusions at 95\% CL in the mass range (70, 110) GeV,
by checking the two dimensional (2D) distributions of any two of the scanned parameters.
From some of the 2D distributions as shown in Fig.~\ref{fig:2D_plots}, correlations between some of the scanned parameters are observed.
$\lambda_2$ tends to have smaller values when the mass of $H_{5}^{0}$ decreases.
While $\lambda_5$ tends to have larger values when $m_{5}$ decreases.
$\lambda_3$ and $\lambda_4$ have tight correlations.
A polynomial fit is performed to each edge of the 2D distributions.
It's observed that all the events fall into the triangular region between the red line with $\lambda_3$ = -1.45$\lambda_4$ + 0.87 and the black line with $\lambda_3$ = -$\lambda_4$, with $\lambda_4$ $\in$ (0.0, 1.55) and $\lambda_3$ $<$ 0.
The strong correlation between $\lambda_3$ and $\lambda_4$ is due to the constraints from the perturbative unitarity of scalar field scattering amplitudes and the bounded-from-below requirement on the scalar potential, as explained in~\cite{Hartling:2014zca}.
Such correlation supports the choosing of $\lambda_3$ =-$\lambda_4$ (= -1.5) for the studies in~\cite{Ismail:2020kqz}.

\end{multicols}
\begin{center}
        \includegraphics[width=0.3\textwidth,height=0.3\textwidth]{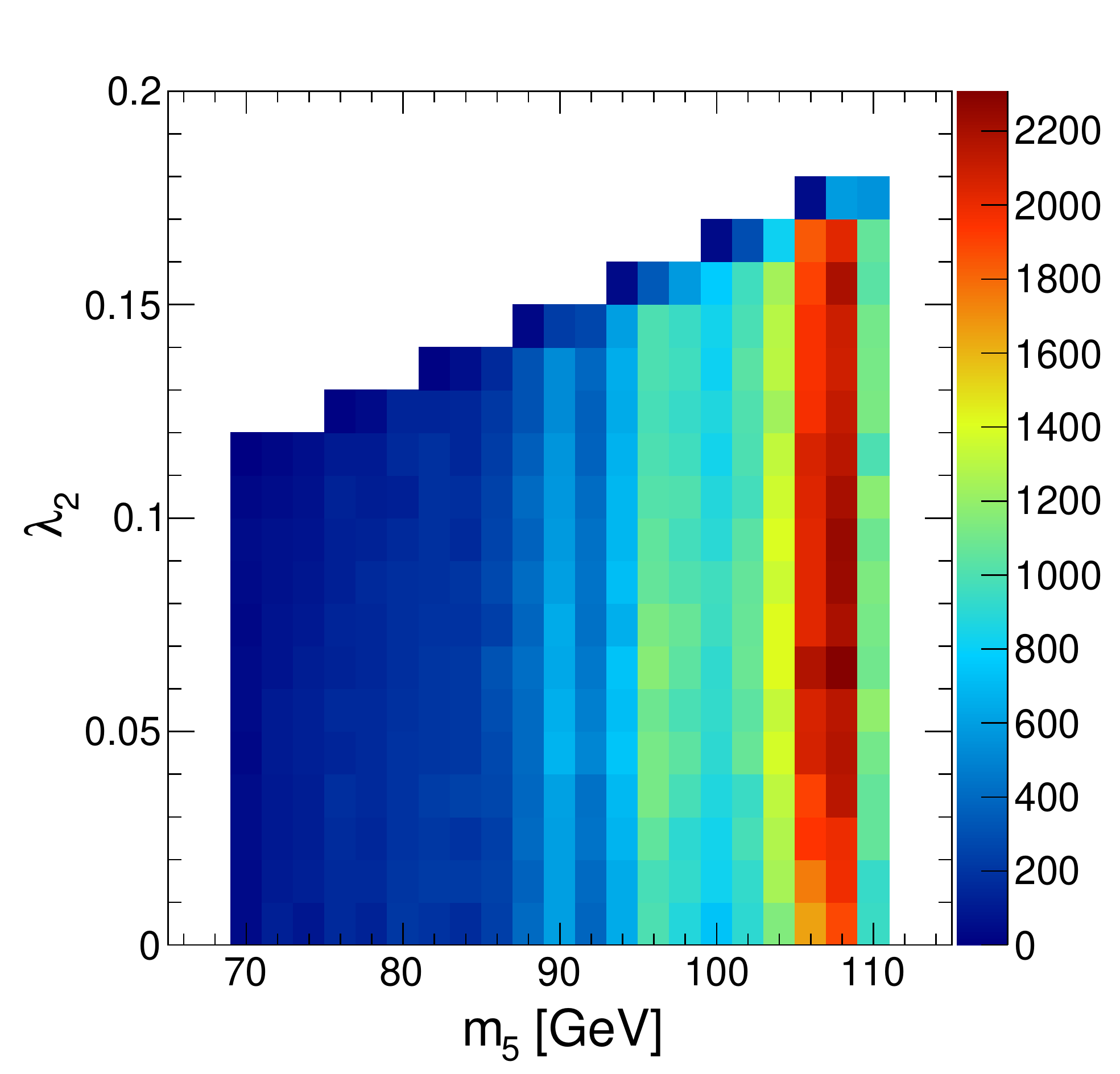}
        \includegraphics[width=0.3\textwidth,height=0.3\textwidth]{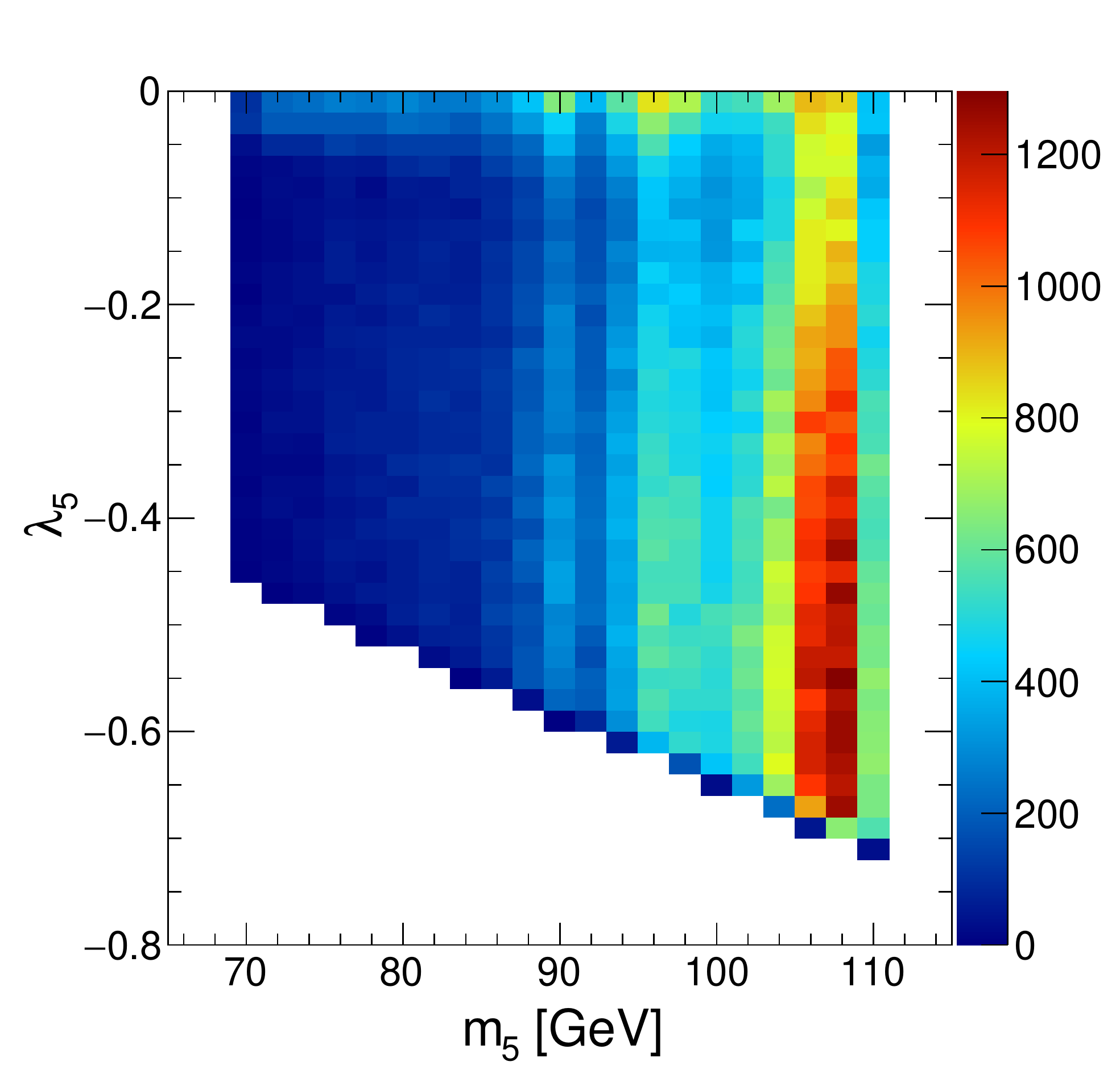}
        \includegraphics[width=0.3\textwidth,height=0.3\textwidth]{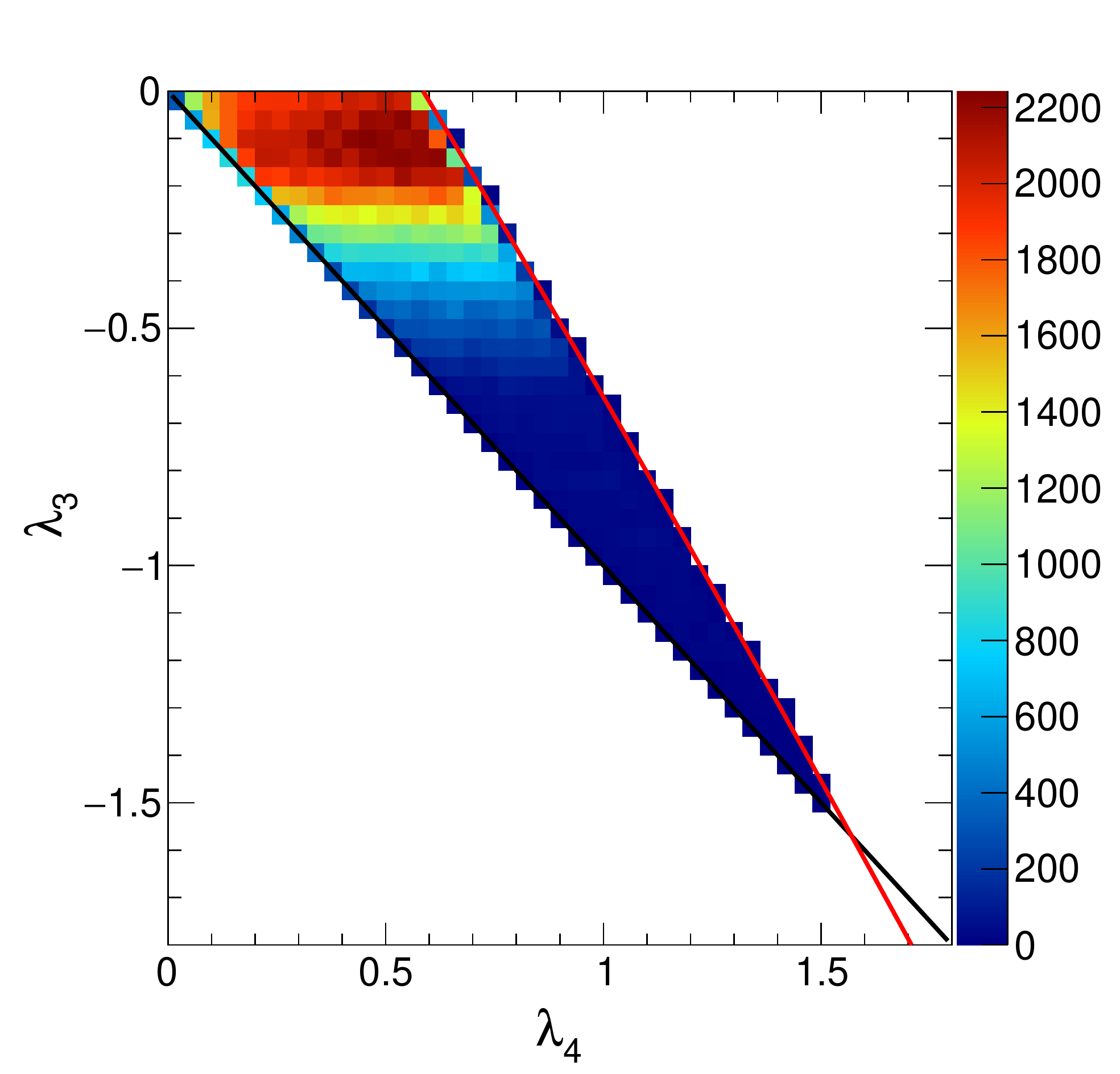}
        \figcaption{\label{fig:2D_plots} (color online) Two dimensional distributions of the GM model parameters, $\lambda_2$ versus ${m_5}$ (left), $\lambda_3$ versus ${m_5}$ (middle), and $\lambda_3$ versus $\lambda_4$ (right) for the points which are not excluded by the CMS observed upper limits in the mass range of [70, 110] GeV.}
\end{center}
\begin{multicols}{2}

\subsection{Discovery potential of other decay channels} \label{sec:OtherScan}

The production rates of the $W^+W^-$, $ZZ$ and $Z\gamma$ decay channels of this low-mass neutral custodial fiveplet scalar in the GM model for 13 TeV pp collisions are also checked,
for the investigation of the discovery potentials of $H_{5}^{0}$ in these channels.
Based on the randomly scanned points after all the constraints as described in subsection 2.2,
the scattering points in Fig.~\ref{fig:Other_channels} show the production rates in pb for $W^+W^-$ decay channel in the left plot, $ZZ$ decay channel in the middle plot and $Z\gamma$ decay channel in the right plot, as functions of ${m_5}$.
The production rates in each decay channel are compared to the predictions (in red lines in the plots) of the SM-like BSM Higgs boson reported by the LHC Higgs Cross Section Working Group~\cite{LHCHiggsCrossSectionWorkingGroup:2016ypw}.
One can expect that there are more chances for the search of low-mass Higgs boson in these decay channels.
As shown in the left plot, for $WW$ decay channel the largest production rate from the GM model prediction at 65 GeV is about 90 times higher than the prediction from the SM-like BSM Higgs boson.
While for $ZZ$ decay channel as shown in the middle plot the largest production rate from the GM model prediction at 65 GeV can reach up to 370 times of the prediction from the SM-like BSM Higgs boson.
Since the cross sections of these two decay channels could be higher to several pb, even after considering the
branching ratios of the cascade decays of $W$ and $Z$ boson it's worth performing the search for this low-mass neutral custodial fiveplet scalar at the LHC with current Run2 data and the near future Run3 data.
In the $Z\gamma$ channel as shown in the right plot, the GM prediction could give larger production rates than the SM-like BSM Higgs boson in the mass range [92,107] GeV.
Due to the lower signal rates of $Z\gamma$ decay and further consideration of the branching fractions of $Z$ decaying into for example leptons, it
is impossible to search for $H_{5}^{0}$ in the $Z\gamma$ channel with LHC Run2 data, but could possible to try with LHC Run3 data and the future High Luminosity LHC (HL-LHC) data.

\end{multicols}
\begin{center}
        \includegraphics[width=0.3\textwidth,height=0.3\textwidth]{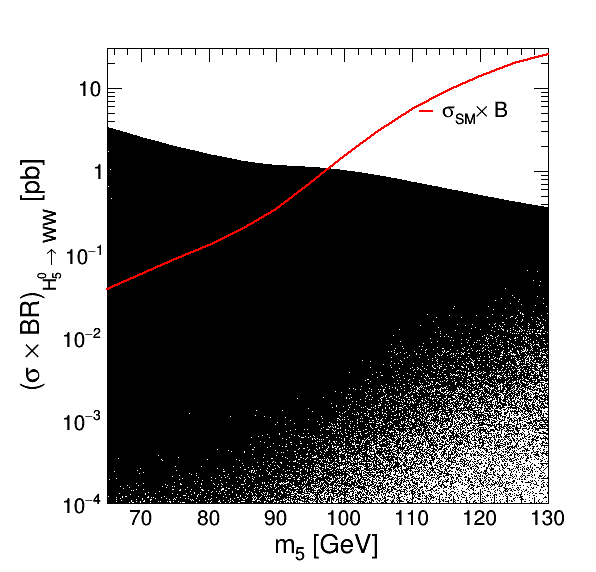}
        \includegraphics[width=0.3\textwidth,height=0.3\textwidth]{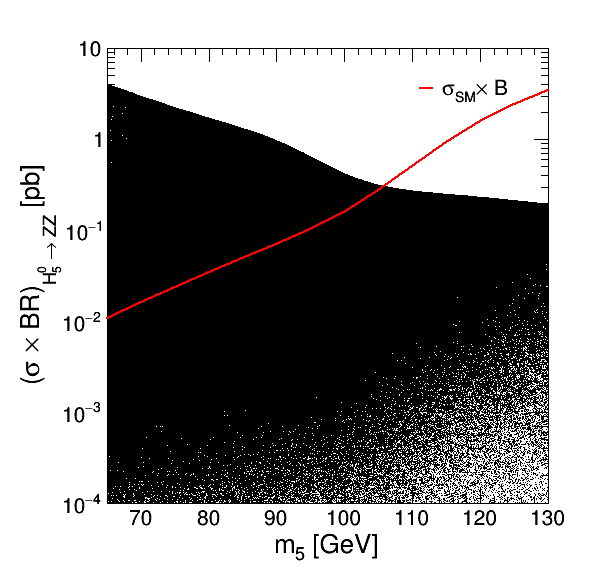}
        \includegraphics[width=0.3\textwidth,height=0.3\textwidth]{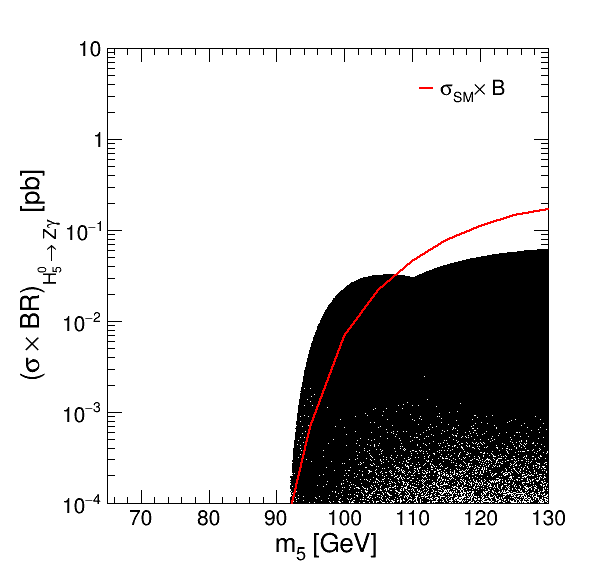}
        \figcaption{\label{fig:Other_channels} (color online) Production rates ($\sigma \times BR$) of $W^+W^-$ (left), $ZZ$ (middle),$Z\gamma$ (right) decay channels of $H_{5}^{0}$ in the GM Model (black points) for 13 TeV pp collisions, with the production rates (red line) at $\sqrt{s}$ = 13 TeV of the SM-like BSM Higgs boson reported by the LHC Higgs Cross Section Working Group~\cite{LHCHiggsCrossSectionWorkingGroup:2016ypw} superimposed for comparisons.}
\end{center}
\begin{multicols}{2}

\section{Conclusions} \label{sec:conclusions}

Search for additional Higgs bosons is one of the most important avenues for probing new physics beyond the Standard Model.
In this paper we explore the possibility of constraining a lighter custodial fiveplet scalar $H_{5}^{0}$ in the Georgi-Machacek model, by restricting the custodial-singlet mass eigenstate $h$ or $H$ to be the LHC observed Higgs boson, after the phenomenological constraints and the constraints from experimental measurements.
To study the phenomenological behavior of a lighter scalar $H_{5}^{0}$, a new set of constraints on the six specific parameters of the GM model as summarized in formula~\ref{eq:gmpara} are proposed to generate events efficiently.
After comparison with the latest results of the search for a lighter Higgs boson with the
diphoton decay channel at 13 TeV by the CMS Collaboration, we conclude that such a lighter scaler $H_{5}^{0}$ has not yet been completely excluded by the LHC experiments.

The CMS observed exclusions at 95\% CL are also used to constrain the possible phase space of the six specific parameters of the GM model.
The tendencies of the GM input parameters are summarized.
For example, for $\lambda_3$ it apparently shows an tendency to higher values populated at around -0.1 for these points which are not excluded by the CMS observed exclusions.
While $\lambda_4$ favors lower values with a peak at around 0.6, showing an opposite behavior compared to $\lambda_3$.
The correlations of the GM input parameters are also checked.
$\lambda_2$  and $\lambda_5$ are some dependent on the $H_{5}^{0}$ mass.
$\lambda_3$ and $\lambda_4$ have very tight correlations.

Finally the discovery potential of other interesting decay channels of this low-mass neutral custodial fiveplet scalar are studied.
For a lighter custodial fiveplet scalar $H_{5}^{0}$, it's worth performing the search for it at the LHC in the $W^+W^-$ and $ZZ$ decay channels, with current Run2 data and the near future Run3 data.
While for the $Z\gamma$ decay channel with lower signal rates, it could be possible to try with LHC Run3 data and the future HL-LHC data.



\vspace{3mm} \emph{The authors would like to thank Yongcheng Wu for helpful discussions and Shulan Zhang for useful suggestions.}

\end{multicols}

\vspace{-1mm} \centerline{\rule{90mm}{0.1pt}} \vspace{2mm}

\input{bib_Paper}

\clearpage

\end{CJK*}
\end{document}

%% file: bib_Paper.tex
\begin{multicols}{2}

\end{multicols}